\documentclass[12pt]{article}
\usepackage{a4}
\usepackage{epsf}
\usepackage{wrapfig}
\usepackage[dvips]{graphicx}
\begin{document}

\begin{center}
{\bfseries ELASTIC PROTON-PROTON AND PROTON-ANTIPROTON SCATTERING:
ANALYSIS OF COMPLETE SET OF HELICITY AMPLITUDES}

\vskip 5mm S.B. Nurushev$^{1}$ and \underline{V.A. Okorokov}$^{2
*}$

\vskip 5mm {\small (1) {\it Institute for High Energy Physics,
142284 Protvino, Moscow Region, Russia}\\
(2) {\it Moscow Engineering Physics Institute (State University),
115409, Moscow, Russia}\\
$^{*}${\it Okorokov@bnl.gov; VAOkorokov@mephi.ru} }
\end{center}

\vskip 5mm
\begin{abstract}
The differential cross-sections are calculated for proton-proton
and proton-anti\-pro\-ton elastic scattering using the
phenomenological model based on the analytic parameterizations for
global scattering parameters (total cross-section and $\rho$ -
pa\-ra\-me\-ter), crossing symmetry and derivative relations. We
confront our model predictions with experimental data in wide
range of energy and momentum transfer. The suggested method may be
useful for PAX Program (GSI) as well as for high-energy
experiments at RHIC and LHC.
\end{abstract}

\vskip 8mm

The elastic proton-proton and proton-antiproton interactions allow
a unique access to a number of fundamental physics observables.
Some important experimental $pp$ and $p\bar{p}$ data are
drastically different in the energy region of $\sqrt{s} < 50$ GeV
and become close each to other at higher energies approaching the
asymptotical expectation. We have proposed earlier two analytical
presentations for full set of helicity amplitudes for $p\bar{p}$
elastic scattering and have made predictions for $t$-dependences
of some spin observables in first presentation
\cite{Okorokov-2006}. In present paper we focus our attention on
predictions for $pp$, $p\bar{p}$ elastic reactions in second
approach.

We use the following analytic parameterization of averaged spin
non-flip amplitude for elastic proton-proton scattering:
\begin{equation}
\Phi_{+}\left(s,t\right) = \sum_{i=1}^{3}A_{i}\delta_{i} \exp
\left(-B_{i}\left(s,t\right)t/2\right),\label{eq:Phi1-pp}
\end{equation}
where $A_{i}$ are free complex constant parameters, the slope
parameters $B_{i}\left(s,t\right)$ are functions of $s$ and $t$,
$\delta_{i}=1,~i=1,3$, and $\delta_{2}=\exp\left(-i\beta\pi
t/2\right)$ describes experimental data in the region of
diffraction deep, $\beta$ - free parameter. We have approximated
the experimental data for slope parameter in order to derive the
analytic energy dependence for $B_{i}\left(s,t\right),~i=1,2$ and
the $B_{3}\left(s,t\right)$ was remained as a free parameter. The
results for different approximations are shown on Fig.~1a, 1b for
energy dependence of slope parameter at low and intermediate $t$
values correspondingly. We choose the following approximation for
the slope parameter:
\begin{equation}
B_{i}\left(s,t\right)=B_{0}^{i}+a_{1}^{i}\left(s/s_{1}\right)^{a_{2}^{i}}
+ 2\alpha_{i}\left[\ln
\left(s/s_{1}\right)\right]^{a_{3}^{i}},~~~i=1,2,\label{eq:B1-B2-pp}
\end{equation}
where $s_{1}=1$ GeV$^{2}$, $a_{3}^{1}=1$ - fixed and
$a_{3}^{2}=1.500 \pm 0.005$. One can see the parameterization
(\ref{eq:B1-B2-pp}) describes all experimental data quite
reasonably for low $t$ domain ($\chi^{2}/ndf$=2.54). The new
experimental data are necessaire at high (RHIC, LHC) energies for
intermediate $t$ values in order to derive more unambiguous energy
dependence of slope parameter. But now the function
(\ref{eq:B1-B2-pp}) approximates this dependence reasonably
($\chi^{2}/ndf$=4.66) and predicts the values of
$B_{2}\left(s,t\right)$ in high energy domain which agree
qualitatively with theoretical expectation $B^{pp}\left(s,t\right)
\simeq B^{p\bar{p}}\left(s,t\right)$ at asymptotic energies.
\begin{figure}[b!]
\begin{center}
\begin{tabular}{cc}
\mbox{\includegraphics[width=7.0cm,height=6.0cm]{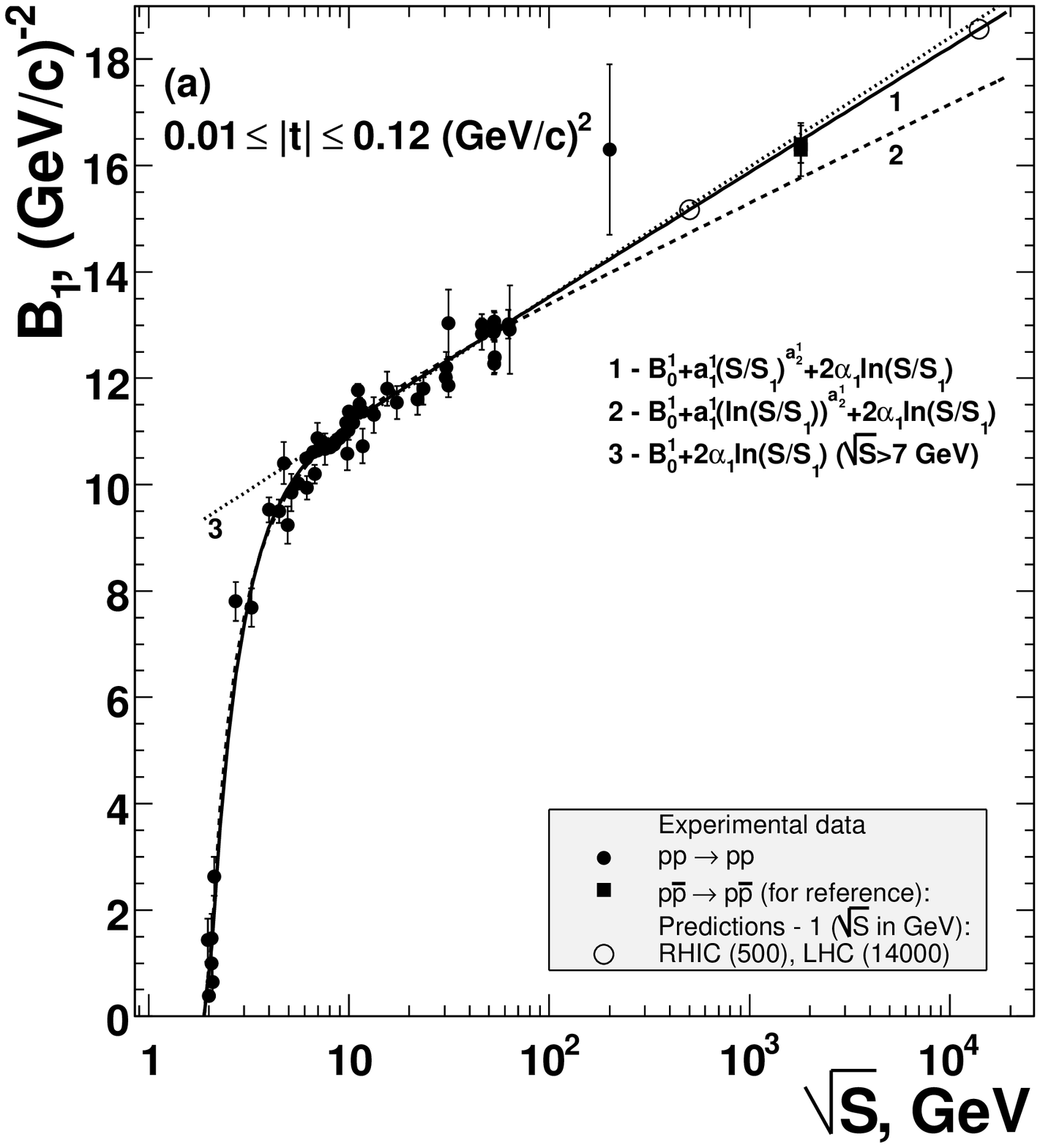}}&
\mbox{\includegraphics[width=7.0cm,height=6.0cm]{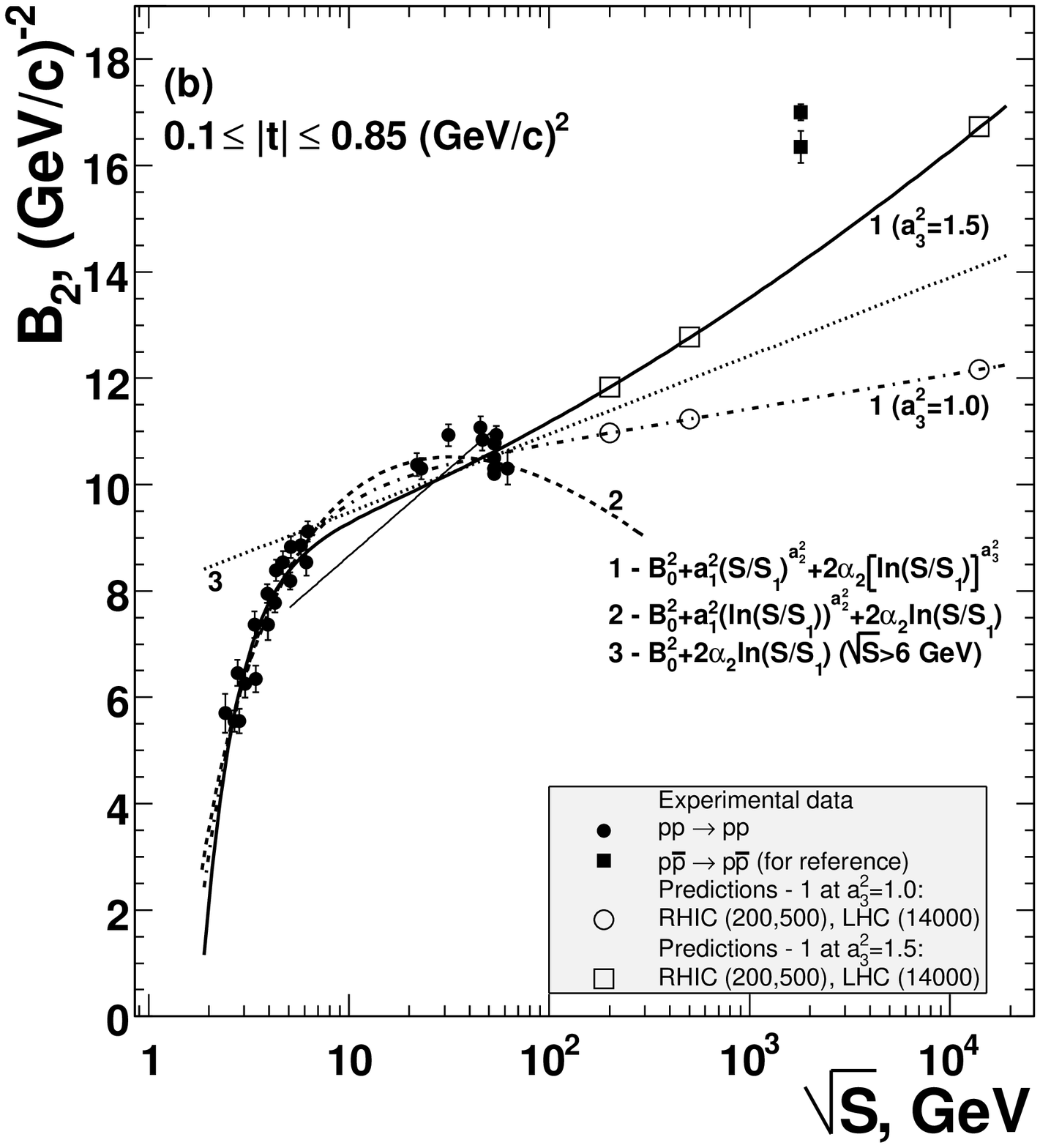}}
\end{tabular}
\end{center}
{\small{\bf Figure 1.} Energy dependence of slope parameter for
low (a) and medium (b) $|t|$ values. The data are drawn from the
Durham Database Group (UK). Thin solid line in (b) is a Regge
model prediction from \cite{Phillips-1973-1}.}\\
\label{SlopeParam-fig1}
\end{figure}
We have followed the standard way \cite{Amos-1988} and have
assumed that the approximation (\ref{eq:Phi1-pp}) describes the
spin non-flip helicity amplitude at $|t|\! \geq \! 0$
(GeV/c)$^{2}$. There are significant data set for total
cross-section $\left(\sigma^{pp}_{tot}\right)$ and
$\rho^{pp}=\left[\Re \Phi_{+}\left(s,t=0\right)\right] / \left[\Im
\Phi_{+}\left(s,t=0\right)\right]$. We choose these two
characteristics for present analysis in order to decrease the
amount of free parameters in (\ref{eq:Phi1-pp}). The PAX project
(GSI), in particular, plans to study the $p\bar{p}$ collisions at
energies $\sqrt{s} \! > \! 3$ GeV. Therefore we have to
investigate this energy domain at least in order to obtain the
reasonable energy dependences of free parameters in spin non-flip
amplitude.

We choose the following parameterizaion for proton-proton total
cross-section:
\begin{equation}
\begin{array}{*{20}c}
\sigma_{tot}^{pp}\left(s\right)=\sum\limits_{j = 1}^{3}\left(\sigma_{tot}^{pp}\right)_{j}, \\
   \begin{array}{l}
\left(\sigma_{tot}^{pp}\right)_{1}=a_{1}\left(\frac{\textstyle
s_{1}}{\textstyle s-4m_{p}^{2}}\right)^{a_{2}};~\left(\sigma
_{tot}^{pp}\right)_{2}=\frac{\textstyle a_{3}}{\textstyle
\xi^{a_{6}-1}}\frac{\textstyle J_{1}(\xi)}{\textstyle
(\xi)},~~\xi=a_{4} \left(s/s_{1}-a_{5}\right); \\
\left(\sigma_{tot}^{pp}\right)_{3}=Z^{pp}+B\ln^{2}\left(s/s_{0}\right)
+Y_{1}^{pp}\left(s_{1}/s\right)^{\eta_{1}}-Y_{2}^{pp}\left(s_{1}/s\right)^{\eta_{2}}. \\
 \end{array}  \\
\end{array} \label{eq:SigmaTot-pp}
\end{equation}
The sum of first two terms is the modification of standard total
cross section parameterization from \cite{PDG-2006} for $\sqrt{s}
\! > \! 5$ GeV.

The different approximations are shown at Fig.2, the fit quality
for (\ref{eq:SigmaTot-pp}) is $\chi^{2}/ndf$=6.95 when using all
available experimental data. As seen the Don\-na\-chie - Landshoff
(DL), Kang - Nikolescu (KN) and standard Particle Data Group (PDG)
parameterizations do not describe the proton-proton total cross
section at low energies. On the other side the suggested approach
describes the $\sigma^{pp}_{tot}$ at qualitative level reasonably
but this fit is still statistically unacceptable. Therefore the
problem of description the low energy data remains open.

Based on the defined analytical parameterization for total
cross-section (\ref{eq:SigmaTot-pp}) one can try to obtain the
corresponding parameterization for $\rho^{pp}$-parameter from
analyticity and the dispersion relations written in the derivative
form. We use the following analytic parameterization for
$\rho^{pp}$-parameter:
\begin{equation}
\rho^{pp}=\frac{\textstyle 1}{\textstyle 2\sigma^{pp}_{tot}}
\left[2\sigma^{pp}_{tot}\Lambda +\sum\limits_{i=1}^{3}
\left(\frac{\textstyle K_{i}}{\textstyle
s}+\pi\delta_{i}\frac{\textstyle
d\left(\sigma^{pp}_{tot}\right)}{\textstyle
d\ln\left(s/s_{1}\right)}\right)
\right],~~\Lambda=\lambda_{1}\frac{J_{1}\left(\lambda_{2}\left(s/s_{1}-\lambda_{3}\right)\right)}
{\textstyle
\left(\lambda_{2}\left(s/s_{1}-\lambda_{3}\right)\right)^{\lambda_{4}}},
\label{eq:rho-pp}
\end{equation}
where the additional term $\Lambda$ describes the low energy data,
the $\sigma^{pp}_{tot}$ are defined above. The first term and
$K_{i}, \delta_{i}, i=1-3$ can be derived from fit of experimental
data. The fit quality for (\ref{eq:rho-pp}) is $\chi^{2}/ndf$=7.8
for all experimental data. For comparison the fit quality is equal
54.4 for PDG parameterization, for example. There are a phase
shift analysis results at energy lower than 5 GeV and we plan to
look at these techniques and improve our description of the
experimental data for low energies.
\begin{wrapfigure}[23]{R}{10.5cm}
\begin{center}
\mbox{\includegraphics[width=8.0cm,height=7.5cm]{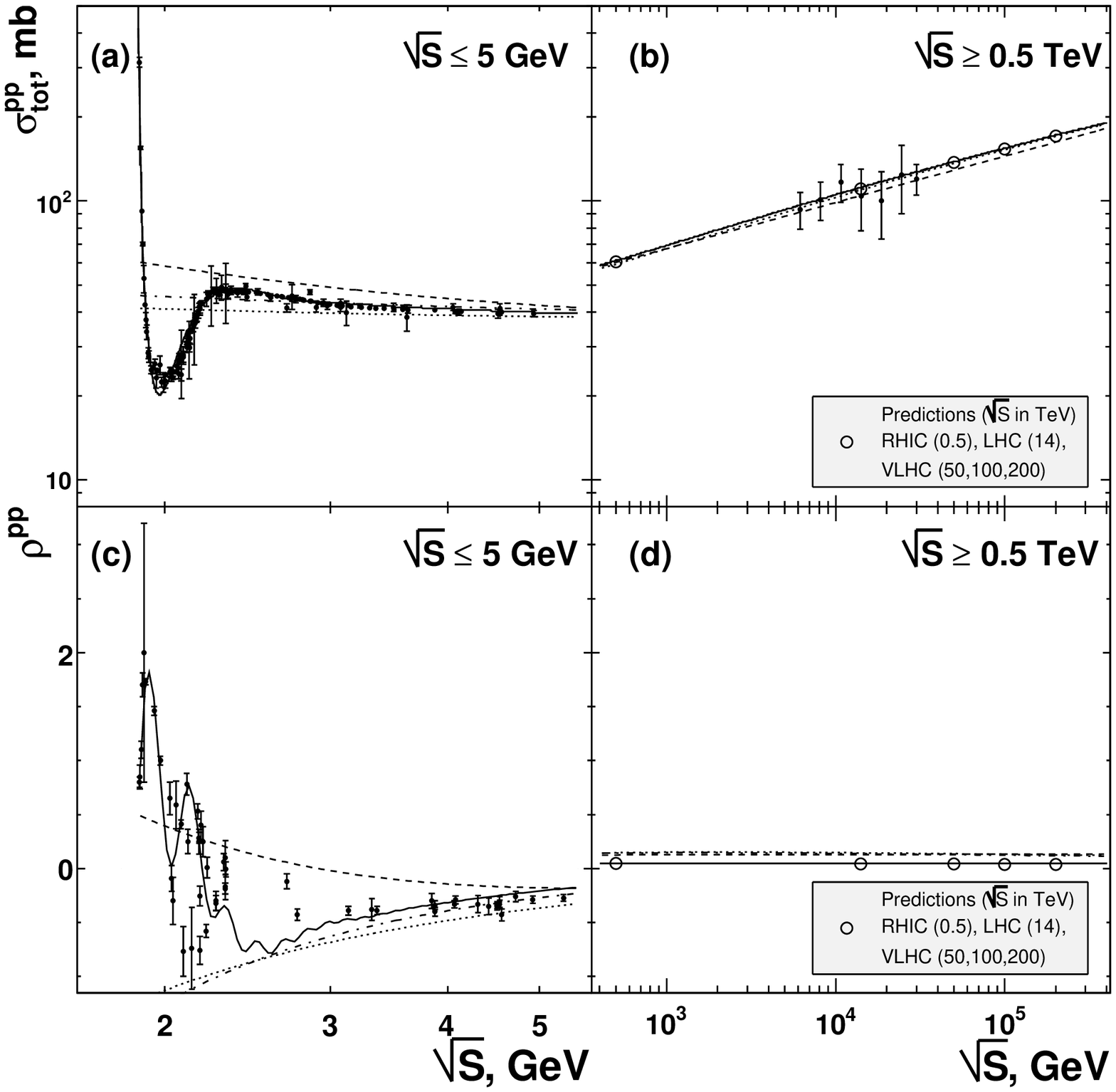}}\\
\end{center}
{\small{\bf Figure 2.} Energy dependences for $\sigma_{tot}^{pp}$
(a,b) and for $\rho^{pp}$ (c,d). Experimental data are from
\cite{PDG-2006}. Solid line is the present work parameterization,
other curves: dashed –- DL, dotted –- KN models
\cite{Avila-2003-1},
dot-dashed –- PDG parameterization \cite{PDG-2006}.}\\
\label{GlobalParameters-fig2}
\end{wrapfigure}

As seen from Fig.~2 the different models predict quite similar
results for $\sigma_{tot}^{pp}$ (Fig.~2b) and for $\rho$ (Fig.~2d)
at high energies, but they valid only above 10 GeV or so. These
models differ at low energies $\sqrt{s} < 5$ GeV dramatically
(Fig.~2a, 2c). Thus we approximated the global scattering
parameters at qualitative level for all available energy domain
and defined $A_{1}$.

The remainding parameters in (\ref{eq:Phi1-pp}) are defined by fit
of experimental proton-proton data for differential cross-section
$d\sigma/dt$, in particular. We have used the method from
\cite{Okorokov-2006} in order to obtain the full set of helicity
amplitudes for proton-proton elastic scattering.

We have considered the data for $pp$ differential  cross-section
in wide energy domain ($\sqrt{s} \! \simeq \! 2-62$ GeV) and for
range of square of transfer momentum $t \! \simeq 10^{-2}-10$
(GeV/c)$^{2}$. Experimental data and corresponding fits are shown
on Fig.~3a for some initial energies. One can see that our
parameterization describes experimental points well at any
energies understudy and up to $t \! \sim \! 9$ (GeV/c)$^{2}$ at
quantitative level. Disagreement between the experimental data and
approximation curves at high $t$ is expected: the high $t$ domain
is described by power dependence inspired pQCD.

We have considered the large set of available experimental data
for $p\bar{p}$ differential cross-section. Analytic curves
contradict with experimental data and some other models (Fig.~3b).
Our approach describes experimental data fairly well at energies
$\sqrt{s} \! \geq \! 19$ GeV at all $t$ values and it's close to
the modified additive quark (mAQ) model. But our approach
contradicts to experimental data and Regge model predictions at
low energies.
\begin{figure}[b!]
\begin{center}
\begin{tabular}{cc}
\mbox{\includegraphics[width=7.0cm,height=6.5cm]{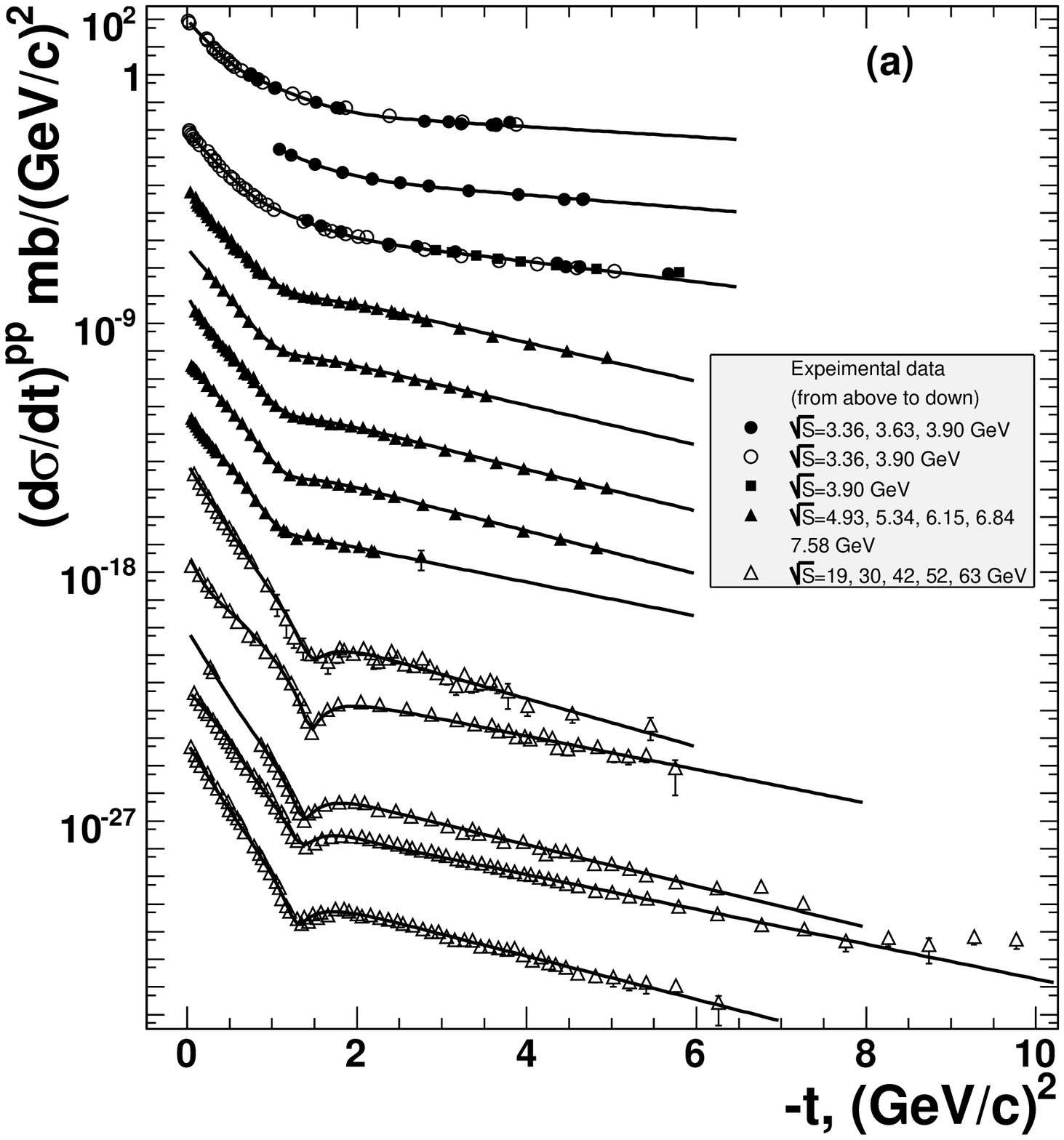}}&
\mbox{\includegraphics[width=7.0cm,height=6.5cm]{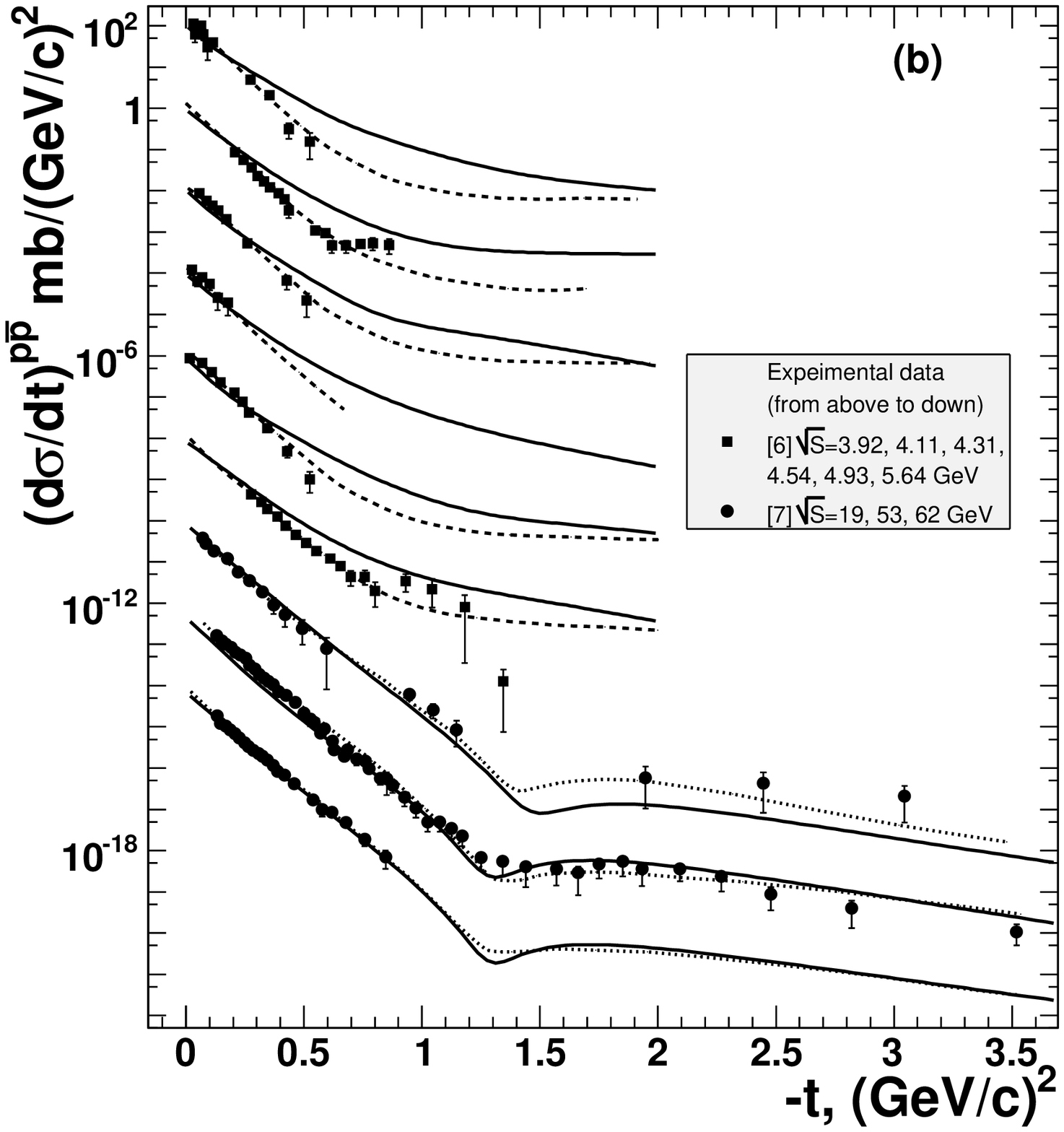}}
\end{tabular}
\end{center}
{\small{\bf Figure 3.} Differential cross sections for elastic
$pp$ (a) and $p\bar{p}$ (b) scattering. A factor $10^{-2}$ between
each successive energy is omitted. Experimental data are from the
Durham Database Group (UK) for $pp$ and from
\cite{Austin-1970,Desgroland-2000} for $p\bar{p}$. Solid lines are
predictions of present work, other curves at (b): dashed –-
Regge-pole \cite{Austin-1970}, dotted –-
mAQ \cite{Desgroland-2000} model prediction.}\\
\label{DiffCrossSection-fig3}
\end{figure}

In summary, the new analytic approach for full set of helicity
amplitudes for elastic $pp$ collisions allows to describe well
proton-proton experimental differential cross section at $\sqrt{s}
\! \simeq \! 2-62$ GeV and up to $t \! \sim \! 9$ (GeV/c)$^{2}$.
Full set of helicity amplitudes for $p\bar{p}$ elastic scattering
is derived based on the known helicity amplitude parameterization
for $pp$ and crossing-symmetry. Analytic approach describes
experimental $p\bar{p}$ data well at $\sqrt{s} \! \geq \! 19$ GeV
and for low and intermediate $t$ value, $t < 1.5$ (GeV/c)$^{2}$.

\end{document}